\documentclass[a4paper]{jpconf}
\usepackage{graphicx}
\usepackage{amssymb}
\usepackage{amsmath}

\begin{document}  

\title{Approaching the precursor nuclei of the third r-process peak with RIBs}

\author{C.~Domingo-Pardo$^{1,*}$, R.~Caballero-Folch$^{2}$, J.~Agramunt$^{1}$,
  A.~Algora$^{1,3}$, A.~Arcones$^{4}$, F.~Ameil$^{4}$, Y.~Ayyad$^{5}$, J.~Benlliure$^{5}$,
  M.~Bowry$^{6}$, F.~Calvi\~no$^{3}$, D.~Cano-Ott$^{7}$, G.~Cort\'es$^{2}$, T.~Davinson$^{8}$,
  I.~Dillmann$^{4,9}$, A.~Estrade$^{4,10}$, A.~Evdokimov$^{4,9}$,
  T.~Faestermann$^{11}$, F.~Farinon$^{4}$, D.~Galaviz$^{12}$,
  A.~Garc\'ia-Rios$^{7}$, H.~Geissel$^{4,9}$, W.~Gelletly$^{6}$,
  R.~Gernh\"auser$^{11}$, M.B.~G\'omez-Hornillos$^{3}$, C.~Guerrero$^{13}$,
  M.~Heil$^{4}$, C.~Hinke$^{11}$, R.~Kn\"obel$^{4}$, I.~Kojouharov$^{4}$,
  J.~Kurcewicz$^{4}$, N.~Kurz$^{4}$, Y.~Litvinov$^{4}$, L.~Maier$^{11}$,
  J.~Marganiec$^{14}$, M.~Marta$^{4,9}$, T.~Mart\'inez$^{7}$, G.~Mart\'inez-Pinedo$^{4}$, B.S.~Meyer$^{21}$,
  F.~Montes$^{15,16}$, I.~Mukha$^{4}$, D.R.~Napoli$^{17}$,
  Ch.~Nociforo$^{4}$, C.~Paradela$^{5}$, S.~Pietri$^{4}$,
  Z.~Podoly\'ak$^{6}$, A.~Prochazka$^{4}$, S.~Rice$^{6}$, A.~Riego$^{2}$,
  B.~Rubio$^{1}$, H.~Schaffner$^{4}$, Ch.~Scheidenberger$^{4,9}$,
  K.~Smith$^{18,19}$, E.~Sokol$^{20}$, K.~Steiger$^{11}$, B.~Sun$^{4}$,
  J.L.~Ta\'in$^{1}$, M.~Takechi$^{4}$, D.~Testov$^{20,22}$, H.~Weick$^{4}$,
  E.~Wilson$^{6}$, J.S.~Winfield$^{4}$, R.~Wood$^{6}$, P.~Woods$^{8}$ and A.~Yeremin$^{20}$}
\address{
$^1$ IFIC, CSIC-University of Valencia, Valencia, Spain\\ 
$^2$ INTE-DFEN, UPC, Barcelona, Spain\\
$^3$ Institute of Nuclear Research of Hungarian Academy of Sciences, Debrecen, Hungary\\
$^4$ GSI Helmholtzzentrum f\"ur Schweionenforschung GmbH, Darmstadt, Germany \\
$^5$ Universidade de Santiago de Compostela, Santiago de Compostela, Spain\\
$^6$ Deptartment of Physics, University of Surrey, Guildford, United Kingdom\\
$^7$ Centro de Investigaciones Energ\'eticas, Medioambientales y Tecnol\'ogicas, Madrid, Spain\\
$^8$ School of Physics and Astronomy, University of Edinburgh, United Kingdom\\
$^9$ Physikalisches Institut, Justus-Liebig Universit\"at Giessen, Germany\\
$^{10}$ St.~Mary's University, Halifax, Nova Scotia, Canada\\
$^{11}$ Physics Department E12, Technische Universit\"at M\"unchen, Garching, Germany\\
$^{12}$ Centro de Fisica Nuclear da Universidade de Lisboa, Lisboa, Portugal\\
$^{13}$ CERN, Geneva, Switzerland\\
$^{14}$ ExtreMe Mater Institute, Darmstadt, Germany\\
$^{15}$ National\! Superconducting\! Cyclotron\! Laboratory,\! Michigan\! State\! University,\! East\!\! Lansing,\! USA\\
$^{16}$ Joint Institute for Nuclear Astrophysics, Michigan State University, East Lansing, USA\\
$^{17}$ Instituto Nazionale di Fisica Nucleare, Laboratori Nazionale di Legnaro, Italy\\
$^{18}$ Department of Physics, University of Notre Dame, South Bend, USA\\
$^{19}$ Joint Institute for Nuclear Astrophysics, University of Notre Dame, South Bend, USA\\
$^{20}$ Flerov Laboratory, Joint Institute for Nuclear Research, Dubna, Russia\\
$^{21}$ Clemson University, USA\\
$^{22}$ Institut de Physique Nucl\'eaire d'Orsay, France\\
}

\ead{domingo@ific.uv.es}

\begin{abstract}
The rapid neutron nucleosynthesis process involves an enormous amount of very exotic neutron-rich nuclei, which represent a theoretical and experimental challenge.
Two of the main decay properties that affect the final abundance distribution the most are half-lives and neutron branching ratios. Using fragmentation of a primary $^{238}$U beam at GSI we were able to measure such properties for several neutron-rich nuclei from $^{208}$Hg to $^{218}$Pb. This contribution provides a short update on the status of the data analysis of this experiment, together with a compilation of the latest results published in this mass region, both experimental and theoretical. The impact of the uncertainties connected with the beta-decay rates and with beta-delayed neutron emission is illustrated on the basis of $r$-process network calculations. In order to obtain a reasonable reproduction of the third $r$-process peak, it is expected that both half-lives and neutron branching ratios are substantially smaller, than those based on FRDM+QRPA, commonly used in $r$-process model calculations. Further measurements around $N\sim126$ are required for a reliable modelling of the underlying nuclear structure, and for performing more realistic $r$-process abundance calculations.
\end{abstract}

\section{Introduction}
The third $r$-process peak at $A\sim195$, the platinum peak, seems particularly sensitive to both the nuclear physics input and the conditions of the stellar environment as can be seen in the detailed sensitivity study reported in Ref.~\cite{Arcones11}. Radioactive Ion Beam (RIB) facilities have enabled the production and measurement of several waiting-point nuclei directly on the $r$-process path from $N=50$ to $N=82$. However all of the nuclei that lead directly to the formation of the third $r$-process abundance peak still remain in the region of the {\it terra incognita}. To a large extent this is due to the very low production cross sections, the limited primary beam intensities available at present RIB facilities and also the challenging experimental conditions of large backgrounds and very low production rates. This contribution focuses on the impact of half-lives and beta-delayed neutrons on the formation of the third $r$-process peak, and how present theoretical models compare with the experimental data available.

\section{The r-process paradigm around N=126}~\label{sec:hl}

The aim of this section is to describe briefly the latest experimental and theoretical results in this mass region, and to explore the impact of the present uncertainties (or discrepancies) on the nucleosynthesis of the $r$ process. On the experimental side, special focus will be made on the latest experiment performed at GSI, thus providing an update on the present status of the data analysis and its future perspective. 

As mentioned above, the nuclei which are directly in the path of the $r$ process along $N=126$, approximately from gadolinium to tantalum, could not be accessed experimentally yet. Nevertheless, one can try to measure neutron-rich nuclei in the higher $Z$ neighbourhood, on both sides of the $N=126$ shell closure, and use such information as a benchmark for theoretical models. In turn, such models can be applied more reliably to extrapolate the decay properties of the $r$-path nuclei.
Three independent experiments in the neutron-rich region around $N=126$ have been carried out recently at the GSI facility for heavy ion research (Germany). The FRS fragment separator~\cite{Geissel92} was used for selection and identification of the ions of interest. The high energy available for primary beams at GSI, of up to 1~GeV/u, has an advantage over other facilities in order to reduce ion identification difficulties related to the charged states of the secondary fragments. Fragmentation of primary $^{238}$U and/or $^{208}$Pb beams on a beryllium target was used for the production of secondary neutron-rich ion beams. These experiments are briefly described below. For details the reader is referred to the cited publications. Only the main results from each experiment are summarised here.

\subsection{$\beta$-decay experiments around N$\sim$126}

The first experiment~\cite{Alvarez09,Alvarez10} used relativistic fragmentation of both $^{238}$U and $^{208}$Pb projectiles for the production of a large number of new neutron-rich nuclei. Several isotopic species were implanted in a stack of four double-sided silicon-strip detectors (DSSSDs)~\cite{Kumar09}, which allowed for the position and time measurement of both implant- and decay-events. $\beta$-Decay half-lives were determined after developing a new numerical method~\cite{Kurtukian08}, which was needed to account for the high and complex background conditions. Half-lives have been published from $^{194}$Re ($N=119$) up to $^{202}$Ir ($N=125$)~\cite{Kurtukian07,Kurtukian09}. From this experiment, another isotopes have been also analysed~\cite{Morales11,Benlliure12}, but they will not be considered in this contribution because they have not been published yet.

New half-lives have been published recently beyond $N=126$, for $^{219}$Bi and $^{211,212,213}$Tl~\cite{Benzoni12}. In the latter experiment, in addition to a stack of three DSSDs for measuring ion-implants and $\beta$-decays, the RISING array~\cite{Wollersheim05} of 105 HPGe detectors was used to detect $\gamma$-rays following the $\beta$-decays. As is nicely illustrated in Ref.~\cite{Benzoni12}, by means of high-resolution $\gamma$-ray spectroscopy it was possible to perform a validation of the numerical technique described in Ref.~\cite{Kurtukian08}, which has been applied in both works~\cite{Kurtukian09,Benzoni12}.

\begin{figure}[!htbp]
\begin{center}
\includegraphics[width=0.8\textwidth]{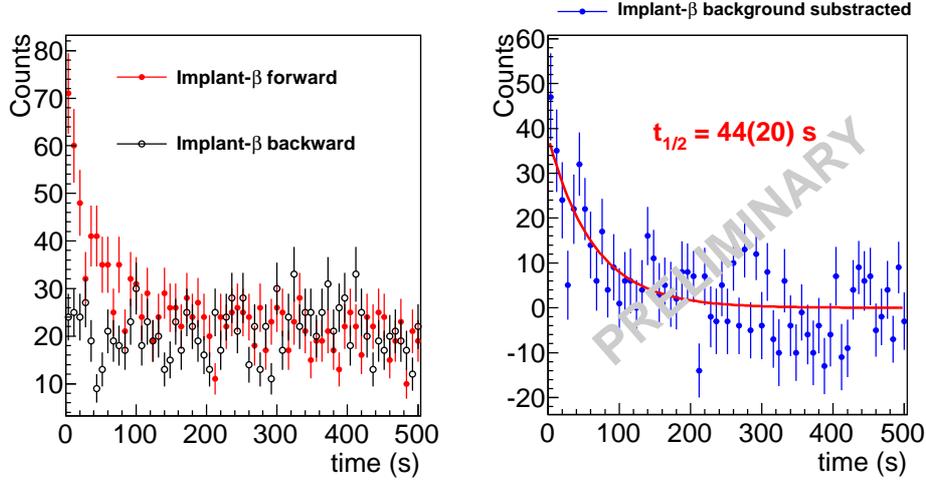}
\caption{\label{fig:212Tl} (Left) $^{212}$Tl Implant-$\beta$ time correlations in the forward direction (solid symbols) and in the backward direction (open symbols) for all decay events within the implant pixel and the 8 neighbouring pixels. All decay events over a broad time window of several times the expected half-life were considered. (Right) Background subtracted correlation spectrum. The quoted half-life value is preliminary.}
\end{center}
\end{figure}

Finally, the latest measurement in this mass region ($N>126$) used a stack of six SSSDs and three DSSSDs~\cite{Steiger09} surrounded by a prototype of the BEta deLayEd Neutron (BELEN) detector~\cite{Gomez11}. The latter allowed for the experimental determination of neutron-branching ratios. Technical details and a few preliminary results have been reported recently in Ref.~\cite{Caballero13}. In summary, the $N>126$ nuclei $^{208-211}$Hg, $^{211-215}$Tl and $^{214-218}$Pb were implanted with sufficient statistics for a reliable analysis of their half-lives. As first approach, the numerical method~\cite{Kurtukian08} was applied to the measured thallium isotopes. As reported in Ref.~\cite{Caballero13}, the preliminary half-lives thus obtained are compatible, within the error bars, with those reported by Benzoni et al.~\cite{Benzoni12} for the common Tl-nuclei. Alternatively, despite the large and complex background environment, we have investigated the possibility of using the more conventional analytical approach based on determining the half-life from implant-beta time correlations. This seems still feasible in those cases with relatively large implant rates (see Fig.~2 in Ref.~\cite{Caballero13}). In the case of e.g. $^{212}$Tl, an ion implant-rate of $\sim$3$\times 10^{-5}$~counts/pixel/s was achieved, being the average rate of $\beta$-like events of about 4$\times 10^{-4}$~counts/pixel/s. Using this nuclide as example for illustration purposes, the corresponding implant-beta time correlations are shown in Fig.~\ref{fig:212Tl}. The correlation area comprises the pixel where the ion was implanted, as well as the eight neighbouring pixels.

In the spectra shown in Fig.~\ref{fig:212Tl} each implanted $^{212}$Tl ion was correlated with all following $\beta$-decay events registered inside the aforementioned correlation area of 9 mm$^2$. The time-window for correlations spanned several times the expected half-life. The background was estimated in a similar way, by performing backward ion-$\beta$ time-correlations over the same DSSSD area. After background subtraction the resulting time-spectrum was adjusted to a simple exponential decay (the daughter nucleus $^{212}$Pb shows a half-life of 10.64~h\cite{ensdf}), as shown in Fig.~\ref{fig:212Tl}. The preliminary value obtained for the half-life of $^{212}$Tl was $t^{^{212}Tl}_{1/2} = 44(20)$~s. This value, although smaller than the one obtained with the numerical method, is still in perfect agreement (within the quoted error bars) with the half-life reported previously. From the 14 isotopic species implanted, we expect that the analytical analysis method can be applied to derive the half-lives of at least half of them, those showing high implant statistics. The remaining cases will be analysed, most probably, by applying the numerical approach~\cite{Kurtukian08}. 

\begin{figure}[!htbp]
\begin{center}
\includegraphics[width=0.49\textwidth]{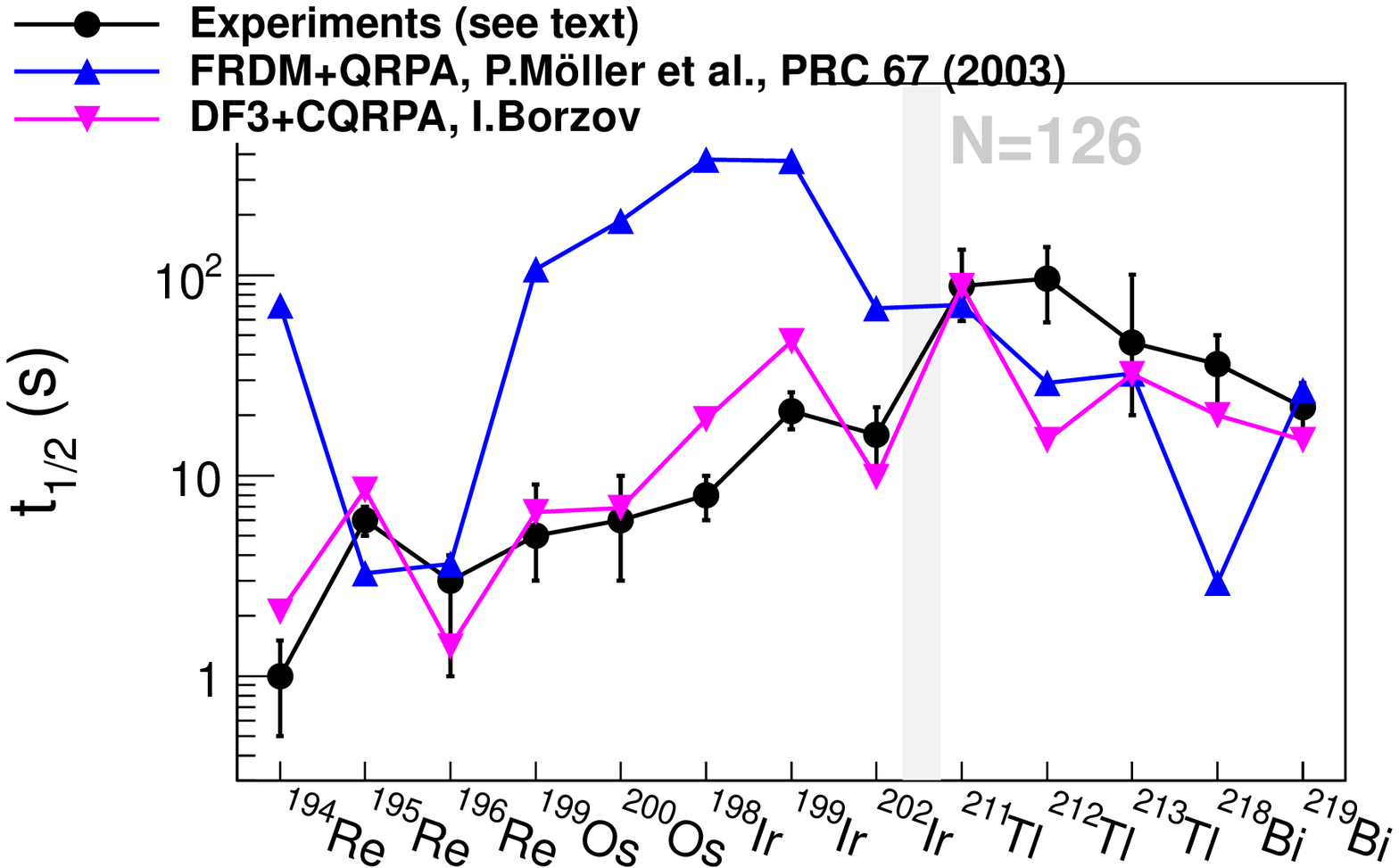}
\hfill
\includegraphics[width=0.49\textwidth]{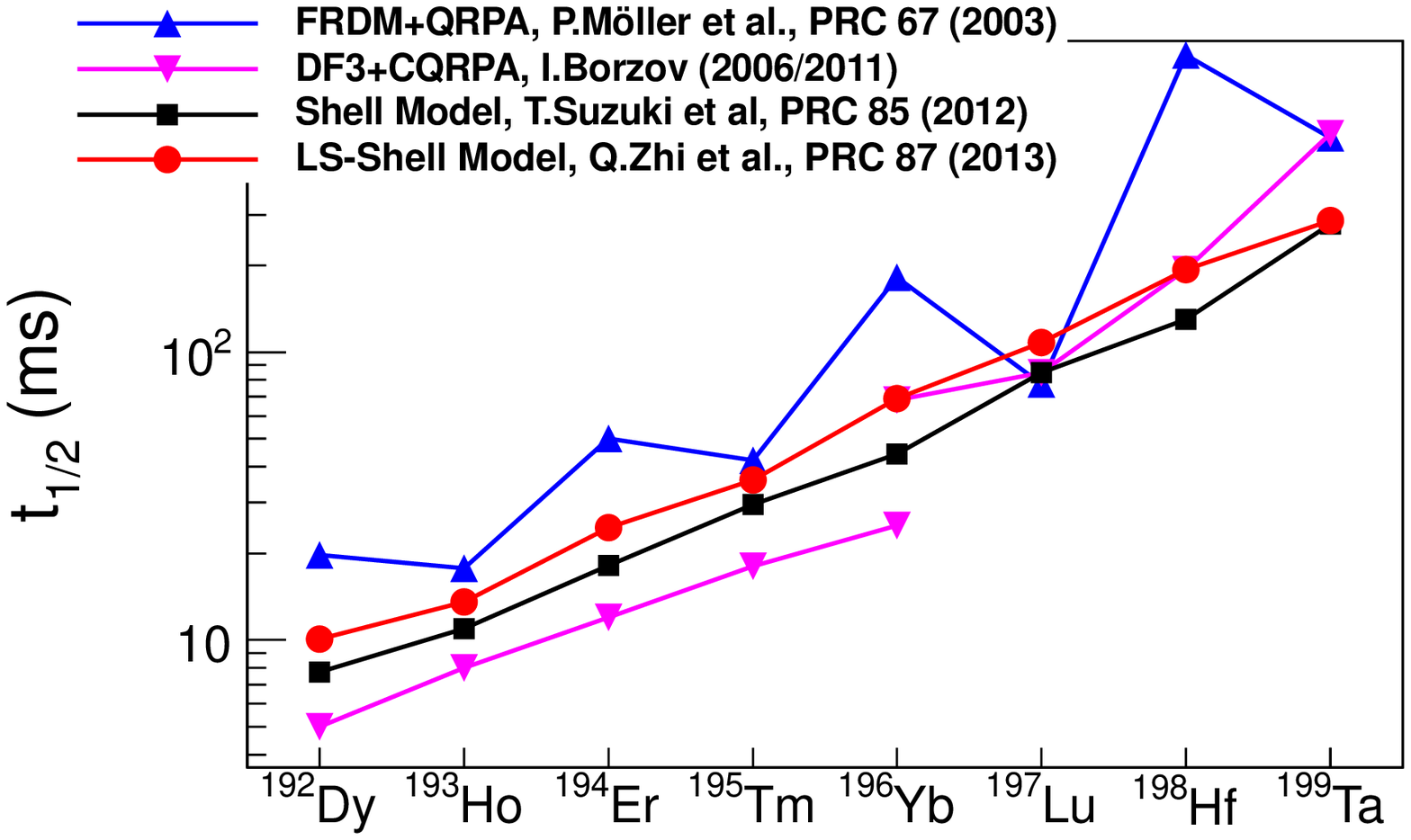}
\caption{\label{fig:hl}(Left) Comparison between theoretical predictions and measurements in the region around $N=126$. Experimental values from $^{194}$Re to $^{202}$Ir are from Ref.~\cite{Kurtukian09}, values between $^{211}$Tl and $^{219}$Bi are from Ref.~\cite{Benzoni12}. (Right) Theoretical values published in the literature for the waiting point nuclei along $N=126$ (adapted from Ref.\cite{Zhi13}).}
\end{center}
\end{figure}

The results for the half-lives determined in these three experiments are displayed in Fig.~\ref{fig:hl}-left. The solid red circle shows the preliminary result discussed above for the half-life of $^{212}$Tl.

\subsection{Theory versus experiment}

Although still far from the $r$-process path, these measurements can be used to judge the reliability of theoretical models in the heavy neutron-rich region around $N \sim 126$. Fig.~\ref{fig:hl}-left shows a comparison between the aforementioned experimental data, and the two main theoretical calculations available in this region. The blue up-pointing triangles correspond to the model of P.~M\"oller~\cite{Moller03}, commonly used in many $r$-process network calculations. This approach is based on the finite-range droplet mass model (FRDM) and uses the quasiparticle random-phase approximation (QRPA) for the Gamow-Teller (GT) part of the $\beta$-strength function, whereas the gross-theory is implemented to account for the first-forbidden (FF) part of the $\beta$-decay. 
The pink down-pointing triangles correspond to the calculations by I.~Borzov using the Fayans energy-density functional (DF3 version) within a continuum QRPA framework~\cite{Borzov03,Borzov11}. The latter approach allows for a self-consistent treatment of both allowed GT- and FF-transitions. Although the contribution of FF-transitions can be neglected at the $N=50$ and $N=82$ shell closures, it is expected that around $N = 126$ they contribute remarkably to the $\beta$-strength distribution and even dominate it above Z$\geq$70, thus providing a remarkable shortening of the half-lives in this mass region~\cite{Borzov03}.


In comparison with the experimental data displayed on Fig.\ref{fig:hl}-left the FRDM+QRPA calculations~\cite{Moller03}, on average overestimate the measured half-lives by a factor of 24 (average ratio of calculation/experiment) in the region $N\leq 126$, and underestimate them by less than 40\% beyond $N=126$ (the average ratio is 0.62). The CQRPA+DF3 model~\cite{Borzov11} agrees better with the experimental data in the region $N\leq 126$, showing an average deviation of only a factor of 2 in that region. The discrepancy becomes again $\lesssim$40\%, in average, beyond the neutron shell closure. 
As already noted in Refs.~\cite{Benlliure12,Benzoni12}, theoretical models seem to overestimate the half-lives before the $N=126$ neutron shell closure, and to underestimate them beyond it. The question, however, is how well these models will perform when one uses them to extrapolate further off stability, into the region of the waiting-point nuclei. To answer this question one can compare the theoretical models among them. In particular shell-model calculations~\cite{Suzuki12,Zhi13}, which became recently available, seem quite helpful given the absence of any experimental information. Such a comparison is shown on the right-hand side of Fig.~\ref{fig:hl}, which has been adapted from Ref.~\cite{Zhi13}. As expected, the agreement of the different models near the neutron shell closure is much better than toward the valley of stability (see Fig.~\ref{fig:hl}-left), owing to the simplified nature of the shell closure. Indeed, comparing the performance of the FRDM+QRPA model against large-scale shell model calculations~\cite{Zhi13} one finds discrepancies, which range from $\sim$30\% up to a factor of 5, with an average deviation of ``only'' a factor of 2. In summary, one can conclude two things. First, the uncertainties on the beta-decay half-lives along the shell closure itself are expected to be less severe than in the $N<126, \,\, Z<82$ mass region covered by the $r$ process during freeze-out. Second, the latest measurements on both sides of $N=126$ provide a very stringent test for theoretical models indeed.

\subsection{R-process nucleosynthesis}
\begin{figure}[h]
\begin{center}
\includegraphics[width=0.49\textwidth]{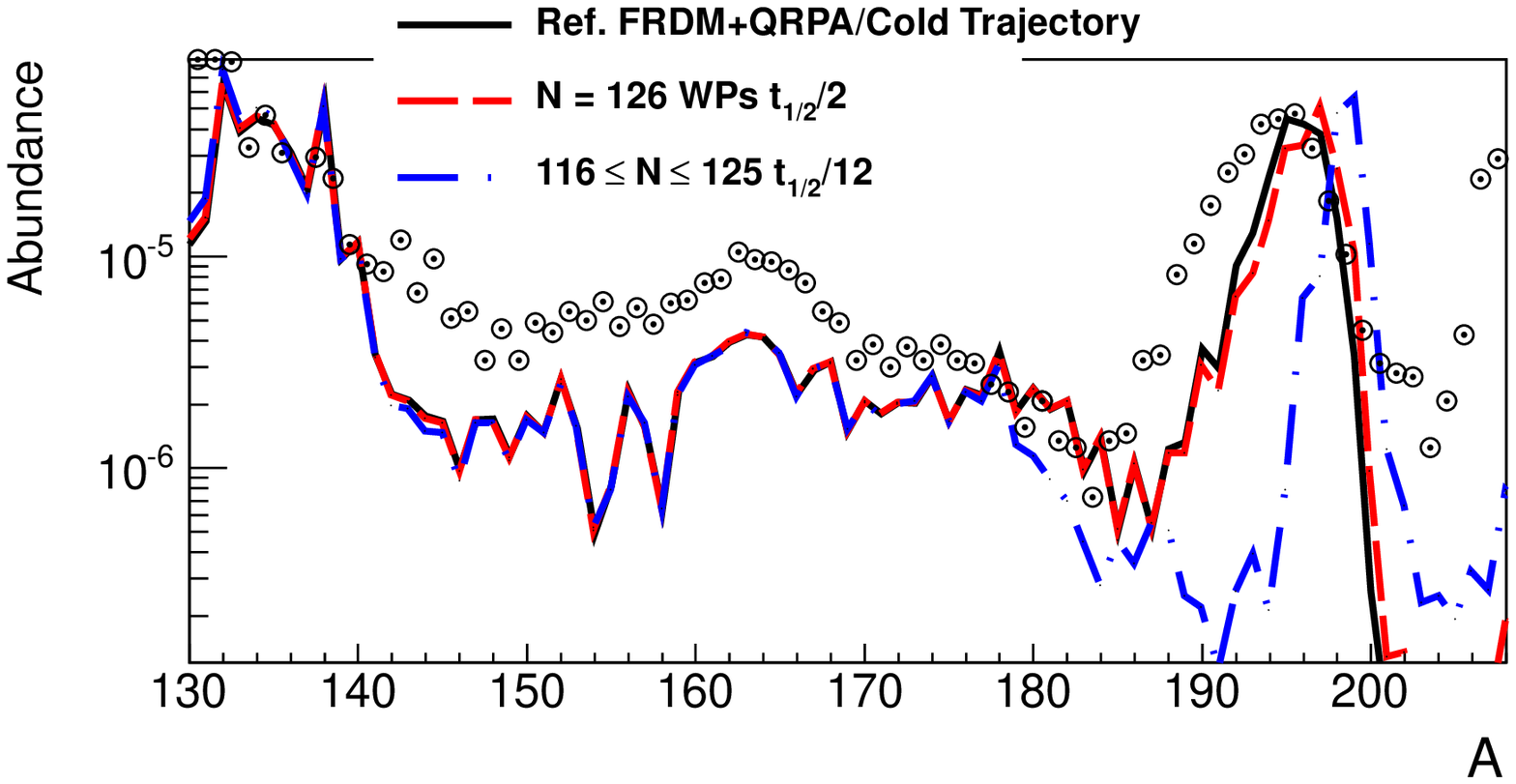}
\includegraphics[width=0.49\textwidth]{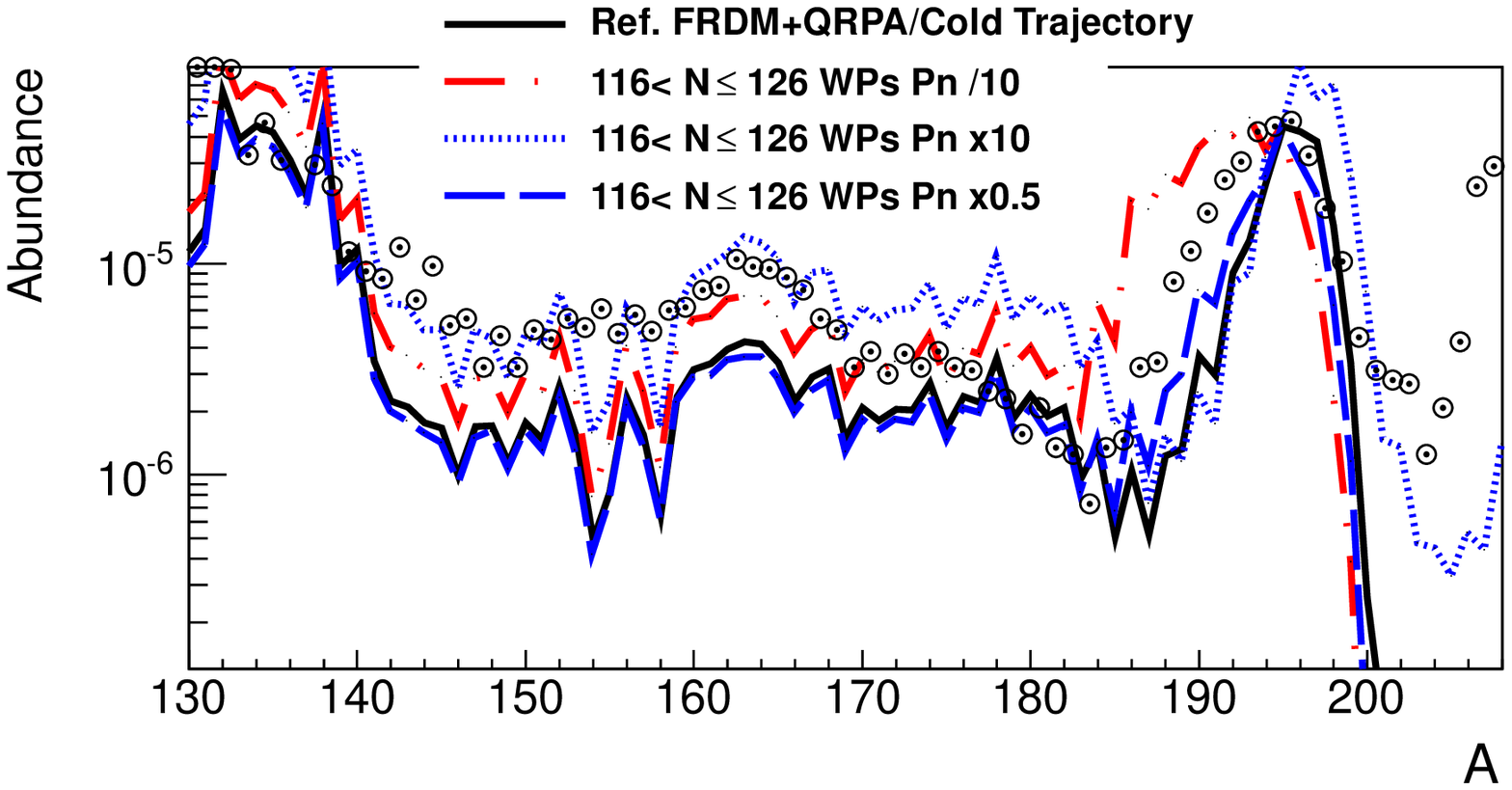}
\caption{\label{fig:yvsa} Abundance versus mass number. Dotted open circles represent solar $r$-process abundances. The solid black curve corresponds to a reference calculation, where the decay rates are essentially based on FRDM+QRPA~\cite{Moller03}. (Left panel) The dashed red line shows the same calculation with a factor of 2 reduction on the half-lives of the $N=126$ waiting point nuclei. The blue dash-dotted line shows the abundances obtained when the half-lives of the nuclei on the left-hand side of the waiting points, from $N=116$ up to $N=125$, were shortened by a factor of 12. See text for details. (Right panel) Dashed-dotted line shows the effect of a reduction by a factor of 10 on the neutron emission probabilities for nuclei in the $116 < N \leq 126$ region. The blue dotted line corresponds to an enhancement, by a factor of 10, of the neutron emission probabilities of the same nuclei. The blue dashed line shows the impact of a reduction by a factor of 2 on the neutron branchings.}
\end{center}
\end{figure}

The impact of the variations in half-lives on the final abundances has been studied in several works (see e.g.~\cite{Borzov08,Suzuki12,Arcones11} and references therein). We have performed new network calculations in order to illustrate the effect of the aforementioned shortening of the half-lives along the $N=126$ waiting-point nuclei, as well as in the $N<126,\,\, Z<82$ neighbourhood. In these calculations the so-called cold trajectory of Ref.~\cite{Arcones11} was used in combination with the single-zone code of \verb+NucNet Tools+~\cite{Meyer12,libnucnet}. Rather standard parameters were used for the electron fraction ($Y_e \approx 0.47$) and the entropy (S$\approx 200 k_B/$nuc). In order to accelerate the calculation of the $r$-process phase of the expansions, we
used Krylov-space iterative matrix solver routines~\cite{Saad03} as implemented in
the freely-available software \verb+Sparskit2+.  
The very sparse network matrix that
must be solved during the $r$-process phase is well-conditioned, so such
iterative solvers converge rapidly, and we found roughly a factor of ten
speed up over the default sparse matrix solver in \verb+NucNet Tools+ when we
used these iterative solvers.
The resulting abundances as a function of the mass number are shown in Fig.~\ref{fig:yvsa}-left. The dotted open circles represent the $r$-process abundances in the solar system. The solid black curve shows a reference calculation performed with the JINA REACLIB library~\cite{Cyburt10}. In the latter database, beta-decay rates are taken from Ref.~\cite{Moller03} (FRDM+QRPA) where no data are available. The calculation provides a reasonable description of the solar $r$-process abundance pattern in the region of the second and third $r$-process peaks. Discrepancies such as the width of the peak or the region between both shell closures are not relevant for the present discussion, where the aim is to illustrate rather qualitatively the impact of reasonable half-life variations in this region. Reducing by a factor of two the FRDM+QRPA half-lives of the $N=126$ waiting-point nuclides, from gadolinium up to tantalum, one obtains the abundances represented in Fig.~\ref{fig:yvsa}-left by the red-dashed line. A small shift towards higher masses is observed, similar to that reported e.g. in Ref.~\cite{Suzuki12}. This seems to be in contradiction with the solar $r$-process abundances, where the maximum appears at a lower mass number around $A\sim195$. Furthermore, as discussed above, the uncertainties on the half-lives of the nuclei in the $N<126$ region seem to be much larger than those at the shell closure itself. Obviously, the uncertainty will depend on the complexity of each nucleus and the difficulties of the model in reproducing the beta-strength distribution and the underlying nuclear structure details. Since it is difficult to assess an uncertainty on each particular nucleus, let us see what is the effect of a variation of the half-lives in that region by a constant factor, which is of the same order-of-magnitude as the discrepancies found between experiment and theory (FRDM+QRPA) discussed above. Assuming an average reduction of ``only'' a factor 12 (the average ratio theory/experiment found above was $\sim$24) in the half-lives of the nuclei on the left-hand side of $N=126$, from $N=116$ up to N$\lesssim$ 125, one obtains a much stronger shift of the third $r$-process peak, towards higher masses, as shown in Fig.~\ref{fig:yvsa}-left by the blue dot-dashed line. Additionally, the third $r$-process peak becomes narrower, an effect that disagrees further with the observed abundance pattern. Thus, one can conclude two things. First, a reduction of the FRDM+QRPA half-lives in the $N\leq 126$ region, which is indeed expected both from experiment and theory, produces a shift of the third $r$-process peak which seems to go in the wrong direction, i.e. toward higher masses. Secondly, with the data currently available, the impact on the final abundances due to half-life uncertainties for nuclei in the region $N<126$, may be even more important than the uncertainties for the waiting-point nuclei themselves.
From the nuclear physics point of view, the discrepancy between shorter half-lives and the position (and width) of the third $r$-process peak, may be counterbalanced by the effect of $\beta$-delayed neutron emission. Indeed, for a given $Q_{\beta}$ value, shorter half-lives usually correlate with a small neutron emission probability. The impact of (smaller) neutron emission probabilities on the final abundances is discussed in Sec.~\ref{sec:neutrons}.

\section{$\beta$-Delayed neutron emission around N$\sim$126}\label{sec:neutrons}

\begin{figure}[h]
\begin{center}
\includegraphics[width=0.45\textwidth]{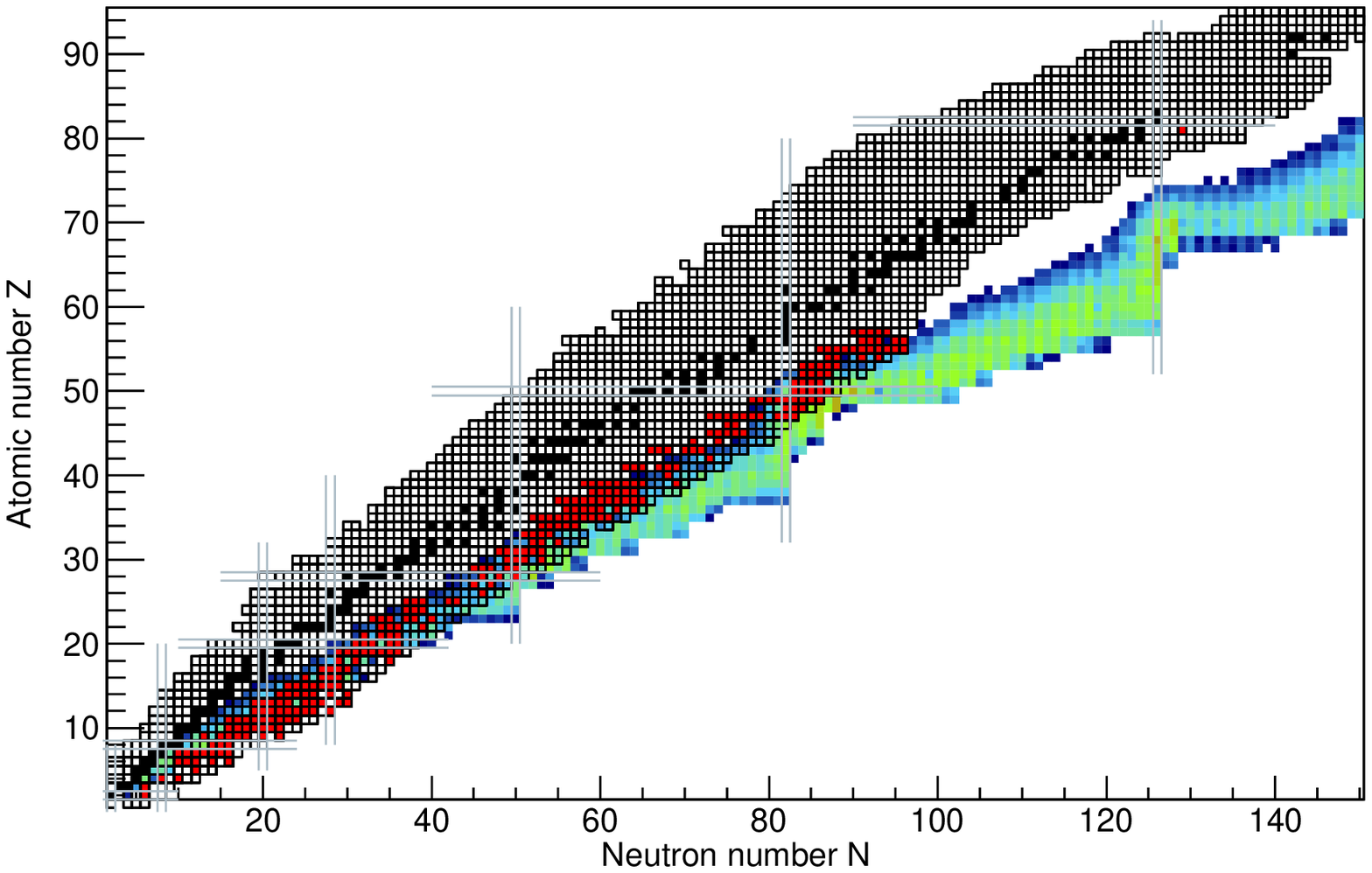}
\includegraphics[width=0.35\textwidth]{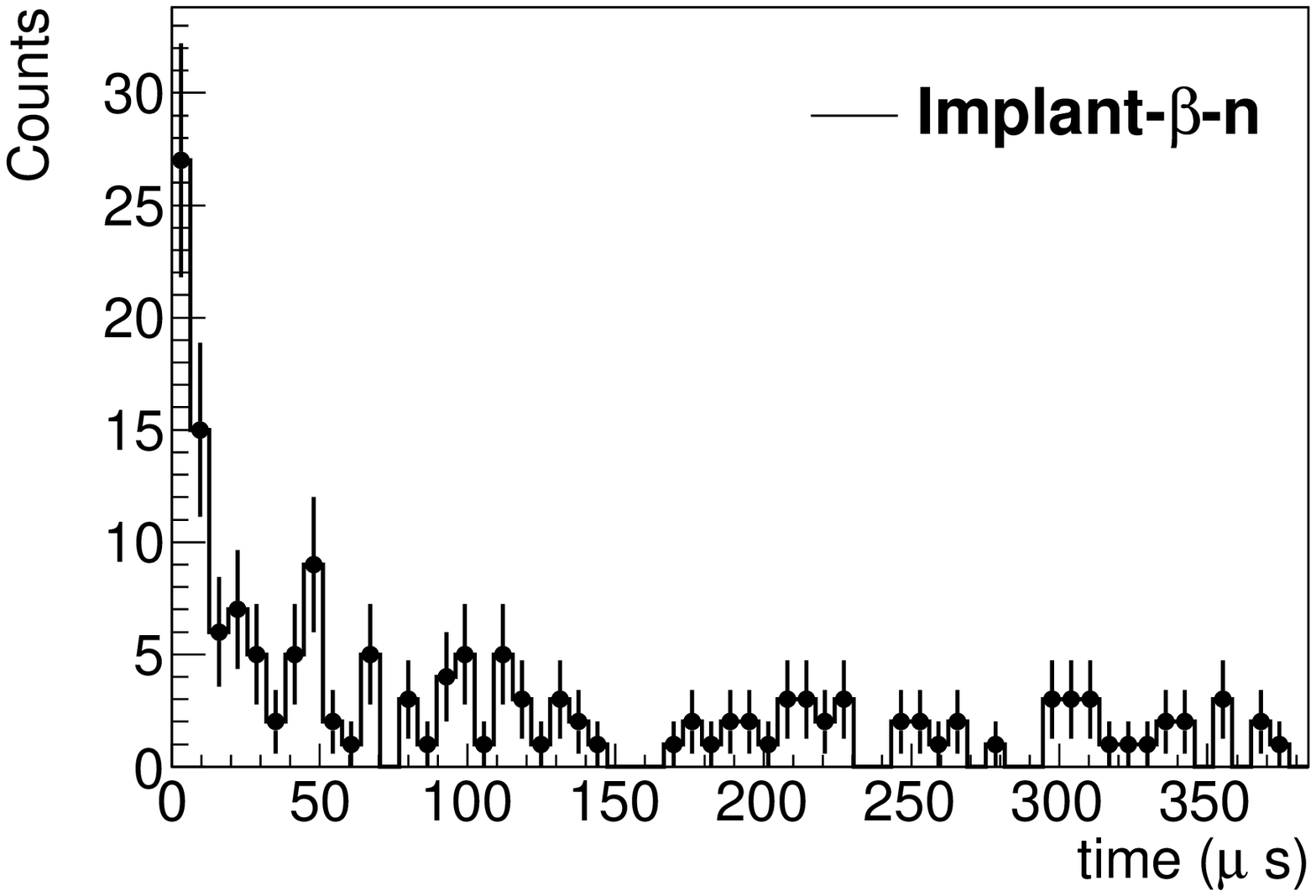}
\caption{\label{fig:pn} (Left) Nuclear chart showing in red nuclei where experimental information on the neutron branching ratio is available~\cite{ensdf}. (Right) Time spectrum showing implant-$\beta$-neutron correlations for $^{212}$Tl.}
\end{center}
\end{figure}

The emission of $\beta$-delayed neutrons plays an important role in the formation of the third $r$-process peak~\cite{Moller03,Arcones11}. However, the situation in terms of experimental information becomes really critical around $N \sim 126$, as demonstrated in Fig.~\ref{fig:pn}-left. 
In the latter figure, nuclei where any experimental information~\cite{ensdf} is known about their beta-delayed neutron emission probability are shown in red. For the sake of comparison, the $r$-process path at some stage of a network calculation is also shown in the same figure.
It is worth emphasising that beyond $A \sim 150$ there is practically no experimental information available. The only data available around $N\sim126$ are on $^{210}$Tl~\cite{Kogan57}.

Motivated by this situation we used the BELEN neutron detector in the last experiment performed at GSI around $N \gtrsim 126$. The detector consisted of 30 proportional counters of $^{3}$He embedded in a polyethylene matrix which served as moderator.  More details about this measurement can be found in Ref.~\cite{Caballero13}. The analysis of the beta-delayed neutron emission probabilities is still ongoing. Following the example shown before for $^{212}$Tl, the spectrum of beta-delayed neutrons registered after the decays of $^{212}$Tl is shown in Fig.~\ref{fig:pn}-right. These are implant-$\beta$-neutron correlations, which could be observed mostly due to the high efficiency of BELEN, which was of $\sim$40\%. Using this kind of correlation analysis, we expect to obtain neutron branching ratios (or upper limits) for all the implanted species.

Since there is no experimental information on neutron-emission probabilities in the mass-region around $N=126$, it is not possible to study reliably the capability of theoretical models for predicting the effect of neutron emission. However, one can again compare the predictions of different models among themselves. One such comparison is shown in Fig.~17 of Ref.~\cite{Zhi13} for the three models described above, FRDM+QRPA~\cite{Moller03}, DF3+CQRPA~\cite{Borzov11} and LSSM~\cite{Zhi13}. For the waiting points with Z$<$69, the neutron emission probabilities predicted by the FRDM+QRPA are, on average, a factor of 3 larger than the values calculated with the LSSM. This ratio becomes $\sim$0.6 for the waiting point nuclei with  Z$\geq$69, mostly due to the contribution of FF transitions. It is not straight-forward to predict, what the uncertainties in the calculated neutron-emission probabilities would be, in particular for the neutron-rich nuclei involved during the freeze-out phase. In this case, variations of the neutron-branching ratios are rather arbitrary, but they can still be used for a qualitative interpretation of the large uncertainties in these quantities.
In this respect, Fig.~\ref{fig:yvsa}-right shows the effect of neutron-branching variations in the mass region $116 < N \leq 126$. Final abundances obtained when the neutron emission probabilities are reduced by a factor of 2 are shown by the blue dashed line. The red dash-dotted line shows the effect of a factor of 10 reduction in the neutron branching ratios. The blue dotted line corresponds to a factor of 10 enhancement of the neutron branching ratios.

In summary, the discrepancy between shorter half-lives and the formation of the third peak discussed in Sec.~\ref{sec:hl} could be compensated, at least to some extent, with a reduction of the neutron emission probabilities. Clearly, new $\beta$-delayed neutron emission measurements are needed around $N \sim 126$ in order to confirm such a hypothesis, and to reduce the still very large contribution of the nuclear physics input uncertainties to the calculated $r$-process abundances.

\section{Summary and outlook}
In summary, both theory and experiment, indicate that half-lives in the $N \sim 126$ region near the $r$-process path should be much smaller than those commonly used in $r$-process model calculations (FRDM+QRPA model~\cite{Moller03}). This, however, would imply a shift of the third $r$-process peak towards higher masses, an effect which is in contradiction with the observed $r$-process abundances. From the nuclear physics input, such a discrepancy could be removed if neutron emission probabilities were also smaller. This seems to be the case, as it has been shown by recent large-scale shell-model calculations~\cite{Zhi13}. In the near future, first experimental values for the neutron emission probabilities of several nuclei beyond $N=126$ will become available from the experiment performed at GSI with the BELEN detector. This will represent a first test of the beta-strength distribution above the neutron separation energy, but clearly new measurements are needed in the neutron-rich heavy-mass region.

\section{Acknowledgments}
We thank the technical staff of the GSI accelerators, the FRS, and the target laboratory for their support during the S410 experiment. This work has been partially supported by the  Spanish Ministry of Economy and Competitivity  under grants FPA2011-24553, FPA 2011-28770-C03-03 and AIC-D-2011-0705. I.D., A.Ev., and M.M. are supported by the Helmholtz association via the Young Investigators group LISA (VH-NG-627).
\section*{References}

\end{document}